\documentclass[10pt,twocolumn]{article}

\usepackage[utf8]{inputenc}
\usepackage{amsmath}
\usepackage{amssymb}
\usepackage{titlesec}
\usepackage{cite}
\usepackage{color}
\usepackage{units}
\usepackage{booktabs}
\usepackage{graphicx}
\usepackage[colorlinks=false]{hyperref}
\usepackage[format=plain,labelfont=it]{caption}
\usepackage[left=1.5cm,right=1.5cm,top=2cm,bottom=2cm]{geometry}

\titleformat{\section}{\centering\normalfont\scshape}{\Roman{section}.}{5pt}{}
\titleformat{\subsection}{\normalfont\it}{\Alph{subsection}.}{5pt}{}
\titleformat{\subsubsection}{\normalfont\it}{\hspace{4mm}\arabic{subsubsection})}{5pt}{}

\newcommand\infoFootnote[1]{%
  \begingroup
  \renewcommand\thefootnote{}\footnote{#1}%
  \addtocounter{footnote}{-1}%
  \endgroup}

\newcommand{\R}{\mathbb{R}} 
\newcommand{\N}{\mathbb{N}}

\newcommand{\Z}{\mathbb{Z}}

\newcommand{\Cc}{\mathcal{C}}

\newcommand{\ub}[1]{\overline{#1}} 
\newcommand{\blind}[1]{\textcolor{white}{#1}}

\newcommand{\xb}{\boldsymbol{x}}
\newcommand{\yb}{\boldsymbol{y}}
\renewcommand{\ub}{\boldsymbol{u}}
\newcommand{\vb}{\boldsymbol{v}}
\newcommand{\Ab}{\boldsymbol{A}}

\newcommand{\zb}{\boldsymbol{z}}

\newcommand{\Bb}{\boldsymbol{B}}
\newcommand{\Cb}{\boldsymbol{C}}
\newcommand{\Db}{\boldsymbol{D}}

\newcommand{\Ib}{\boldsymbol{I}}
\newcommand{\Fb}{\boldsymbol{F}}
\newcommand{\Gb}{\boldsymbol{G}}

\newcommand{\Tb}{\boldsymbol{T}}
\newcommand{\Rb}{\boldsymbol{R}}
\newcommand{\Pb}{\boldsymbol{P}}

\newcommand{\wb}{\boldsymbol{w}}
\newcommand{\eb}{\boldsymbol{e}}

\newcommand{\zerob}{\boldsymbol{0}}

\newcommand{\modu}{\mathrm{mod}}

\newcommand{\Enc}{\mathtt{Enc}}
\newcommand{\Dec}{\mathtt{Dec}}

\title{\vspace{-2mm}\bf Encrypted dynamic control with 
unlimited operating time via FIR filters}
\author{Nils Schl\"uter, Matthias Neuhaus, and Moritz Schulze Darup\vspace{2mm}}
\date{}

\begin{document}

\maketitle

\textbf{\textit{Abstract}.} {\bf Encrypted control enables confidential controller evaluations in cloud-based or networked control systems. From a technical point of view, an encrypted controller is a modified control algorithm that is capable of computing encrypted control actions based on encrypted system outputs. Unsurprisingly, encrypted implementations of controllers using, e.g., homomorphic cryptosystems entail new design challenges. For instance, in order to avoid overflow or high computational loads, only a finite number of operations should be carried out on encrypted data. Clearly, this guideline is hard to satisfy for dynamic controllers due to their recursive nature. To enable an unlimited operating time, existing implementations thus rely on external ``refreshments'' of the controller state, internal refreshments using bootstrapping, or recurring controller resets.

We show in this paper that simple FIR filter-based controllers allow to overcome many drawbacks of the existing approaches. In fact, since FIR filters consider only a finite amount of the most recent input data, the recursion issue is immediately solved and controller refreshments or resets are no longer required. Moreover, well-designed FIR filters are often less complex than and equally effective as IIR controllers.}
\infoFootnote{N. Schl\"uter, M. Neuhaus, and M. Schulze Darup are with the \href{https://rcs.mb.tu-dortmund.de/}{Control and~Cyber-physical Systems Group}, Faculty of Mechanical Engineering, TU Dortmund University, Germany. E-mails:  \href{mailto:nils.schlueter@tu-dortmund.de}{\{nils.schlueter, matthias.neuhaus, moritz.schulzedarup\}@tu-dortmund.de}. \vspace{0.5mm}}
\infoFootnote{\hspace{-1.5mm}$^\ast$This paper is a \textbf{preprint} of a contribution to the 19th European Control Conference 2021.}

\section{Introduction}
\label{sec:intro}

Modern control systems tend to be more networked and distributed. Prominent examples come along with industry~4.0, smart grids, building automation, robot swarms, or intelligent transportation systems. While these systems offer exciting features, they also involve privacy and security concerns. A proper controller design for a networked system should take these concerns into account and guarantee confidentiality and integrity of the involved process data.

One step in this direction are encrypted controllers that ensure confidentiality of plant data, controller parameters, and control actions throughout the control loop(s) (see~\cite{darup2020encrypted} for an overview). The defining feature of such controllers is their encrypted evaluation. A key technology for the realization of encrypted control is homomorphic encryption (HE) that allows computations on encrypted data (see, e.g., \cite{Paillier1999,Gentry2010,fan2012somewhat,Cheon2017_CKKS}). However, encrypting control algorithms with homomorphic cryptosystems is a non-trivial task that requires tailored controller reformulations.

Such reformulations and subsequent encryptions have been proposed for various control schemes. For instance, encrypted realizations of static linear state and output feedback can be found in~\cite{Kogiso2015,Farokhi2017}. More complex model predictive control schemes were presented in~\cite{Alexandru2018,SchulzeDarup2018_LCSS}. Recently, encrypted versions of linear dynamic controllers of the form
\begin{subequations}
\label{eq:normalController}
\begin{align}
\label{eq:normalControllera}
\xb(k+1)&=\Ab \xb(k)+\Bb \yb(k), \qquad \xb(0):=\xb_0 \\
\label{eq:normalControllerb}
\ub(k)&=\Cb \xb(k)+\Db \yb(k)\\[-6mm]
\nonumber
\end{align}
\end{subequations}
gained some interest in the literature  \cite{Cheon2018need,Murguia2020,Kim2016,kim2019dynamic}.
Here, ${\xb(k) \in \R^l}$ is the controller state at time step $k\in\N$,  ${\yb(k) \in \R^l}$ refers to feedback from the plant (such  as, e.g., the plant’s output), and ${\ub(k) \in \R^m}$ denotes the resulting control action. Controllers of type~\eqref{eq:normalController} are interesting since many popular control schemes, such as linear-quadratic Gaussian (LQG) control, Luenberger observers combined with linear state feedback, or PID control, can be expressed in this form. Thus, a well-designed encrypted implementation of~\eqref{eq:normalController} can have tremendous impact on the practicality of encrypted control. Unfortunately, despite the simple controller structure, such an implementation is not straightforward.  In this context, the main challenge is to cope with the recursive nature of~\eqref{eq:normalController}.  In fact, the recursive evaluations of the controller states might lead  to (undetectable) overflows during the encrypted computations and hence to false control actions. Without additional means, this effect limits the operating time of encrypted dynamic controllers.

Several approaches for solving this issue have been proposed in the literature \cite{Cheon2018need,Murguia2020,Kim2016,kim2019dynamic}. As detailed in Section~\ref{sec:summary}, these approaches either require recurring ``refreshments'' or resets of the encrypted controller, which increases the evaluation effort or decreases the control performance. In this paper, we show that simple FIR filter-based approximations of dynamic controllers allow avoiding these drawbacks. In fact, FIR filters can be easily encrypted without relying on any refreshments or resets and, if well-designed, they provide good control performance. We address these advantages in more detail below and illustrate the efficiency of our novel approach with a numerical \mbox{benchmark}.

The remainder of this paper is organized as follows. In Section~\ref{sec:encryptedcontrol}, we provide basics on HE and explain why an unlimited encrypted operation of~\eqref{eq:normalController} is difficult. Next, we briefly summarize existing solutions to this problem Section~\ref{sec:summary}. Section~\ref{sec:FIR} is dedicated to FIR approximations for dynamic controllers. Afterwards, benefits for FIR filter-based encrypted control are pointed out in Section~\ref{sec:benefits} and illustrated with a numerical example in Section~\ref{sec:benchmark}. Finally, we state conclusions and an outlook in Section~\ref{sec:outlook}.

\textit{Notation}. We denote the sets of real, integer, and natural numbers by $\R$, $\Z$, and $\N$, respectively.
By $\lfloor\cdot \rfloor$, $\lceil\cdot \rceil$, and $\lfloor\cdot \rceil$, we denote the floor function, the ceiling function, and rounding to the nearest integer, respectively.

\section{Fundamentals of encrypted dynamic control}
\label{sec:encryptedcontrol}

Existing encrypted implementations of linear dynamic controllers (as specified in Section~\ref{sec:summary}) all build on HE schemes. Hence, we begin with a brief overview on HE. Applying these cryptosystems (typically) requires an integer-based reformulation of~\eqref{eq:normalController}, which we address next. This reformulation will also reveal why an unlimited operation time of an encrypted dynamic controller is problematic.

\subsection{Homomorphic encryption}
\label{subsec:HE}

HE schemes allow mathematical operations to be carried out on encrypted data (see, e.g,~\cite{Paillier1999,Gentry2010,fan2012somewhat,Cheon2017_CKKS,ElGamal1985}). More precisely, let $z_1$ and $z_2$ be two arbitrary numbers in the cryptosystem's message space and denote the encryption and decryption procedure by ``$\Enc$'' and ``$\Dec$'', respectively. Then, we call a cryptosystem multiplicatively homomorphic if there exists an operation ``$\otimes$'' that supports encrypted multiplications according to
\begin{equation}
\label{eq:multiplicativelyHomomorphic}
z_1 \, z_2 = \Dec \left( \Enc(z_1) \otimes \Enc(z_2) \right).
\end{equation}
Analogously, cryptosystems are called additively homomorphic if an operation ``$\oplus$'' exists such that 
\begin{equation}
\label{eq:additivelyHomomorphic}
z_1 + z_2 = \Dec \left( \Enc(z_1) \oplus \Enc(z_2) \right).
\end{equation} 
Popular realizations of such partially HE schemes are due to ElGamal~\cite{ElGamal1985} and Paillier \cite{Paillier1999}, respectively.

\begin{figure}[tp]
	\centering
	\includegraphics[trim=4.6cm 21.3cm 6cm 1.8cm, clip=true,width=.9\linewidth]{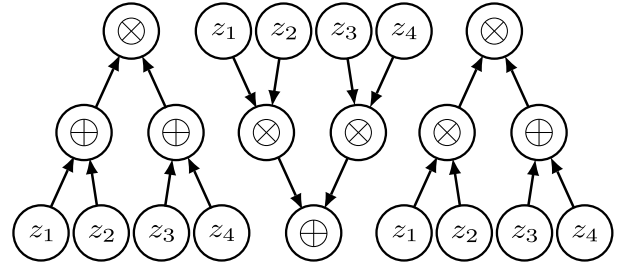}
	\caption{Arithmetic circuits corresponding to (from left to right) the three expressions $(z_1+z_2)(z_3+z_4)$, $(z_1 z_2)+(z_3 z_4)$, and $(z_1 z_2)(z_3+z_4)$. The multiplicative depth of the circuits are 1, 1, and 2, respectively.}
	\label{fig:circuits}
	\vspace{-4mm}
\end{figure}

There also exist cryptosystems that support both homomorphisms~\eqref{eq:multiplicativelyHomomorphic} and~\eqref{eq:additivelyHomomorphic}. These cryptosystems are quite powerful since encrypted multiplications and additions allow, in principle, the secure implementation of arbitrary functions (in terms of boolean or arithmetic circuits). However, only computationally expensive fully HE schemes (as, e.g., \cite{Gentry2010}) support an unlimited number of homomorphic multiplications and additions. In contrast, in leveled HE schemes such as \cite{fan2012somewhat} or \cite{Cheon2017_CKKS}, the number of operations is limited. The supported ``level'' refers to the maximally allowed multiplicative depth of the arithmetic (or boolean) circuit to be encrypted. In this context, the multiplicative depth refers to the maximum number of multiplications on any path from an input node to an output node of the arithmetic circuit. Figure~\ref{fig:circuits} illustrates the notion based on three examples. In order to provide a correct decryption, the number of levels should be (at least) equal to the multiplicative depth of the circuit to be encrypted.

Some leveled homomorphic cryptosystems can be transformed into fully homomorphic ones by (recurrently) using ``bootstrapping''. In a nutshell, bootstrapping evaluates the decryption algorithm homomorphically at high computational cost with help of the encrypted secret key. The result is a ciphertext on an improved level, which allows for further encrypted computations. 

To better understand the concept of levels, we briefly note that most leveled HE schemes build on variants of the learning with errors (LWE) problem introduced in \cite{regev2009lattices}. In these schemes, small errors are added during encryption
in order to ensure security. Unfortunately, these errors grow during homomorphic operations (especially during multiplications) and, at some point, would impair the decryption. Hence, the ciphertext is either decrypted before this happens or bootstrapping is needed.

\subsection{Integer-based controller reformulation}
\label{subsec:integerReformulation}
The message space of most homomorphic cryptosystems is a finite set of integers, e.g.,
$$
\Z_q:=\left\{-\left\lfloor \frac{q}{2} \right\rfloor,\dots,\left\lceil\frac{q}{2} \right\rceil-1\right\},
$$
where $q\in \N_{>1}$ refers to the number of elements. Hence, an encrypted implementation of~\eqref{eq:normalController} requires to  map all involved quantities to  $\Z_q$. For a given number $x \in \R$, this mapping can be realized as follows. We pick a suitable scaling factor $s\in\R_{\geq 1}$, multiply $x$ with this factor, and round the result to the nearest integer, i.e., $z:=  \lfloor s x \rceil$. Three observations are important in this context. First, we obtain $z \in \Z_q$ for a sufficiently large choice of $q$. Second, an approximation of $x$ can be recovered from $z$ by evaluating $z/s$, where the (absolute) approximation error is limited by $1/(2s)$. Third, an analog procedure can be applied to matrices yielding $ \lfloor s \Ab \rceil\in\Z_q^{n \times n}$. Computations with the resulting integers are straightforward, presupposed that additions are carried out based on identical scaling factors. In other words, the relation
$$
\frac{1}{s_1 s_3} \left( \lfloor s_1 x_1 \rceil + \lfloor s_2 x_2 \rceil \right) \lfloor s_3 x_3 \rceil
\approx (x_1+x_2) x_3 
$$
is meaningful if and only if $s_1=s_2$. Following this simple rule, it is easy to see that
\begin{subequations}
\label{eq:integerController}
\begin{align}
\label{eq:integerControllera}
\zb(k+1)&=\lfloor s_1\Ab \rceil \zb(k)+\lfloor s_2\Bb \rceil \lfloor s_1^{k+1} s_5 \yb(k) \rceil\\
\label{eq:integerControllerb}
\vb(k)&=\lfloor s_3 \Cb \rceil \zb(k)+ \lfloor s_4\Db \rceil \lfloor  s_1^{k+1} s_5\yb(k) \rceil\\[-5.5mm]
\nonumber
\end{align}
\end{subequations}
with $\zb(0):= \lfloor s_0 \xb_0 \rceil$ is an integer-based reformulation of~\eqref{eq:normalController} if $s_0=s_2s_5$ and $s_1 s_4=s_2 s_3$ hold. This reformulation does, in principle, support an encrypted implementation based on the homomorphisms~\eqref{eq:multiplicativelyHomomorphic} and~\eqref{eq:additivelyHomomorphic}. In fact, approximations of the control inputs and states could be recovered from
 \begin{equation}
\label{eq:recoveries}
\ub(k) \approx \frac{1}{s_1^{k+1} s_4 s_5} \vb(k) \quad \text{and} \quad \xb(k) \approx \frac{1}{s_0 s_1^k} \zb(k),
\end{equation}
respectively. Now, a simple choice that satisfies the two scaling conditions is $s_0=s_1=s_2=s_3=s_4=s$ and $s_5=1$. At this point, it is important to note that $s_1$ plays a special role. In fact, the integer variables $\zb(k)$, $\lfloor s_1^{k+1} s_5 \yb(k) \rceil$, and $\vb(k)$ all accumulate this scaling factor over time. Hence, if $s_1>1$ and $q$ is finite, these integers will (most likely) leave $\Z_q$ for some $k \in \N$. An encrypted implementation of~\eqref{eq:integerController} would then result in an (undetected) overflow and false control actions. Since we typically require $s_1 \gg 1 $ for small approximation errors, ensuring an unlimited operation time of encrypted dynamic controllers is not trivial.

\vspace{-2mm}
\section{Summary of existing approaches}
\label{sec:summary}

Several approaches that deal with the problem of limited operation times have been proposed in the literature. We briefly summarize these approaches in order to prepare and distinguish our novel procedure.

\subsection{External refresh of controller state}
\label{subsec:external}

Although an unlimited encrypted execution of~\eqref{eq:integerController} is demanding, it is usually straightforward to choose the cardinality $q$ such that an overflow can be excluded for $T\in \N_{\geq 1}$ time steps. A simple approach to circumvent future overflows, which has, e.g., been applied in \cite{SchulzeDarup2020_IFAC}, then is as follows. At time step $k=T-1$, the cloud not only returns $\Enc(\vb(T-1))$ to the plant but also $\Enc(\zb(T))$, i.e., the encrypted controller state for the next step. Both quantities are decrypted at the actuator and approximations for $\ub(T-1)$ and $\xb(T)$ are computed according to~\eqref{eq:recoveries}. While $\ub(T-1)$ is applied to the system, $\xb(T)$ is forwarded to the sensor. Now, at time step $T$, the sensor measures $\yb(T)$ and sends $\Enc(s_1 s_5\yb(T))$ together with $\Enc(s_0\xb(T))$ to the cloud. There, $\Enc(s_0 \xb(T))$ is used to reinitialize the encrypted dynamic controller. As a consequence, the accumulation of $s_1$ is reset and the controller can run for another $T$ time steps before the procedure is repeated. Clearly, a major drawback of this method is the communication overhead. In fact, every $T$ time steps, $n$ additional ciphertexts have to be transmitted from and to the cloud, which is a significant increase. Furthermore, the external refresh is vulnerable to communication issues like latency or packet losses.

\subsection{Internal refresh of controller state} 
\label{subsec:internal}
The communication overhead can be avoided with fully HE. In fact, with help of bootstrapping, it is also possible to ``refresh'' the encrypted controller state within the cloud. However, as pointed out in Section~\ref{subsec:HE}, this is computationally expensive. In \cite{Kim2016}, Kim et al. deal with this issue by orchestrating three controllers to bridge the waiting time when one of them is bootstrapped. Clearly, the parallel evaluation of multiple controllers results in an additional overall computational load.

\subsection{Periodic reset of controller state} 
\label{subsec:reset}

The approaches above either result in additional communication or computational load.
A simple but effective modification allows overcoming these drawbacks. In fact, as recently proposed in \cite{Murguia2020}, a periodic reset of the controller state removes the need for refreshments. According to \cite[Eq.~(2)]{Murguia2020}, the resulting controller then has the form
\begin{align}
\nonumber
\xb_r(k+1)&= \left\{ \begin{array}{ll}
\!\xb_{0} & \text{if} \,\, k+1 \; \mathrm{mod} \; T = 0, \\
\!\Ab \xb_r(k) + \Bb \yb(k) & \text{otherwise} \end{array} \right. \\
\label{eq:resetController}
\ub_r(k)&=\Cb \xb_r(k)+\Db \yb(k).
\end{align}
However, the abrupt resets of $\xb_r(k)$ directly affect the plant's inputs through $\ub_r(k)$, which might degrade the control performance. Nevertheless, closed-loop stability can be ensured using the tailored controller design in \cite[Sect.~III.A]{Murguia2020}.

\subsection{Integer controller state matrix}
\label{subsec:integer}

As shown in Section~\ref{subsec:integerReformulation}, the (un)limited operation issue is crucially linked to the scaling factor $s_1$. As apparent from~\eqref{eq:integerController} and~\eqref{eq:recoveries}, an unlimited operation time of the encrypted controller can easily be guaranteed for the special case~${s_1=1}$ since the accumulation is affectless then\footnote{For $s_1=1$, the conditions on the remaining scaling factors are, e.g., satisfied for $s_2=s_3=s_5=s$ and $s_0=s_4=s^2$.}. Obviously, choosing $s_1=1$ is only reasonable if $\Ab$ is already integer (or close to being so). In general, designing linear dynamic controllers with $\Ab\in\Z^{n \times n}$ is non-trivial. However, as pointed out in~\cite[Sect.~IV]{Cheon2018need}, FIR filters naturally lead to integer $\Ab$ and, for PID controllers, the design problem is considerably simplified. Apart from the generally difficult design, we recently showed in \cite{Schlueter202x} that integer $\Ab$ also involve stability issues. In fact, an $\Ab\in\Z^{n \times n}$ is Schur stable if and only if all eigenvalues are~$0$ (which corresponds to a FIR filter). Thinking (again) about the effects of packet losses and latency, unstable controllers seem unsuitable for networked systems. Hence, exploiting integer $\Ab$ has to be done with care (not only) in encrypted~control.

\subsection{Integer matrix and external refresh of control action}
\label{subsec:hybrid}

A hybrid approach, that combines the techniques from Sections~\ref{subsec:external} and~\ref{subsec:integer}, has recently been presented in the preprint~\cite{kim2019dynamic}.
The authors propose to ``extend'' \eqref{eq:normalControllera} by $\Rb \ub(k) - \Rb \ub(k)=\zerob$ and to substitute \eqref{eq:normalControllerb} in the second term. As a result, one obtains the modified controller
\begin{align*}
\xb(k+1)&=\left(\Ab-\Rb \Cb\right) \xb(k)+\left(\Bb-\Rb\Db\right) \yb(k)+\Rb \ub(k) \\
\ub(k)&=\Cb \xb(k)+\Db \yb(k).
\end{align*}
Now, $\Rb \in \R^{n \times m}$ and a transformation matrix $\Tb \in \R^{n\times n}$ can be chosen such that $\Tb \left(\Ab-\Rb\Cb\right) \Tb^{-1}$ becomes integer (presupposed the pair $(\Ab,\Cb)$ is observable) without suffering from the restrictions mentioned in Section~\ref{subsec:integer}. However, an encrypted implementation of the modified controller requires an external (or internal) refresh of the control action $\ub(k)$ in every time step. While this is a significant drawback, it might be advantageous over the approach in Section~\ref{subsec:external} if $m \ll n$.

\section{Controller approximations with FIR filters}
\label{sec:FIR}

From the existing approaches, the ones in Sections \ref{subsec:external} and \ref{subsec:internal} are dominated by cryptographic tools. In fact, in these approaches, the controller is not alternated apart from quantizations during the integer reformulation. In contrast, Sections~\ref{subsec:reset}, \ref{subsec:integer} and~\ref{subsec:hybrid}, take a more control-oriented viewpoint and the controllers are specifically designed to fit into the framework of HE. In this paper, we follow the latter methodology and present another control-related approach. More precisely, we show that FIR filter-based controller approximations are well-suited for encrypted dynamic control. To this end, we combine the observation that FIR filters are beneficial for encrypted control (see Sect.~\ref{subsec:integer}) with methods for the approximation of infinite impulse response (IIR) dynamical controllers.

In this paper, a FIR filter-based controller is specified as
\begin{equation}
\label{eq:uFIR}
\ub_f(k)= \sum_{j=0}^{N} \Fb_j \yb(k-j),
\end{equation}
where $N \in \N$ denotes the filter order. Under the assumption that a dynamic controller of the form~\eqref{eq:normalController} is given, the first step towards an encrypted realization with a FIR filter is the identification of suitable filter matrices $\Fb_0,\dots,\Fb_{N} \in \R^{m \times l}$. There exist numerous methods to solve this task in the literature (see, e.g., \cite{diniz2010digital}). Here, we consider two schemes that turn out to be  useful and illustrative for our approach.

\subsection{Window-based}
\label{subsec:window}

One of the most basic methods to design a FIR filter is based on windowing. To apply this technique here, we first recall the well-known explicit formula
\begin{equation}
\label{eq:uExact}
\ub(k)= \Cb \Ab^{k} \xb_0 + \sum_{j=0}^{k-1} \Cb \Ab^j \Bb\yb(k-1-j)  + \Db \yb(k)
\end{equation}
for the control actions resulting from~\eqref{eq:normalController}. Obviously, for ${k\geq N}$, \eqref{eq:uExact} can be rewritten as
$$
\ub(k)= \Cb \Ab^{N} \xb(k-N) + \sum_{j=0}^{N-1} \Cb \Ab^j \Bb\yb(k-1-j)  + \Db \yb(k).
$$
Under the assumption that $\Ab$ is Schur stable, $\Cb \Ab^{N} \xb(k-N)$ is small for sufficiently large $N$. Hence, by neglecting this term, we find $\ub(k)\approx \ub_f(k)$ for the choice
\begin{equation}
\label{eq:FjWindow}
\Fb_0 :=\Db \,\,\,\, \text{and} \,\,\,\, \Fb_j := \Cb \Ab^{j-1} \Bb \quad \forall \,\, j\in \{1,\dots,N\},
\end{equation}
which corresponds to a (rectangular) window of length $N$. As illustrated with the numerical example in Section~\ref{sec:benchmark}, this FIR filter parametrization already yields reasonable approximation quality for moderate $N$ if $\Ab$ is Schur stable. More complex variants (such as \cite{kootsookos1992nehari}) even allow accounting for the term $\Cb \Ab^{N} \xb(k-N)$ more precisely. These techniques are, however, beyond the scope of this paper.

\begin{figure}[tp]
	\centering
	\includegraphics[trim=1.4cm 26.5cm 11.5cm 0cm, clip=true,width=\linewidth]{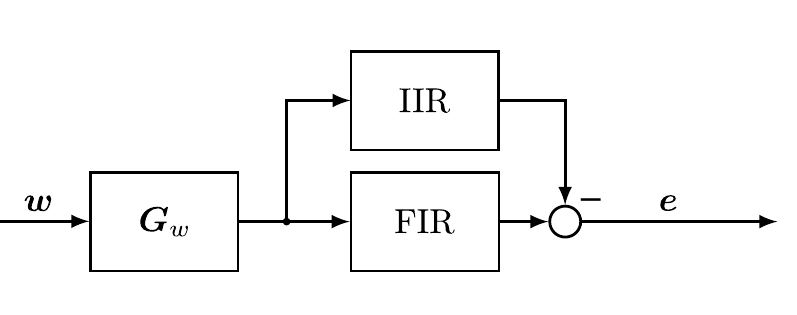}
	\caption{Error dynamics considered in \cite{yamamoto} for an optimal FIR filter-based approximation of IIR filters. A user-defined weighting of the exogenous input $\wb$ can be specified by the transfer function $\Gb_w$.} 
	\vspace{-4mm}
\label{fig:errordynamics}
\end{figure}

\subsection{Optimization-based}
Since we are looking for as accurate as possible controller approximations, it is natural to include optimization-based methods. One such method, that deals with $H_\infty$-optimal FIR filter-based approximations, is presented in \cite{yamamoto}.
Applying the method requires a state-space representation of the FIR filter such as, e.g., 
\begin{align*}
\Ab_f&:=\begin{pmatrix}
    \zerob & \cdots  &  \zerob  &  \zerob \\
    \Ib_l &    & \zerob  & \zerob  \\
     & \ddots & & \vdots \\
    \zerob &   &  \Ib_l & \zerob
    \end{pmatrix} 
    &\Bb_{f}&:=\begin{pmatrix}
    \Ib_l \\
    \zerob \\
    \vdots \\
    \zerob 
    \end{pmatrix} \\
    \Cb_f&:=\begin{pmatrix}
    \Fb_{1} &   \cdots & \Fb_{N-1}  &  \Fb_N
    \end{pmatrix}& \Db_{f}&:=\Fb_0.
\end{align*}
Next, the error dynamics in Figure~\ref{fig:errordynamics} are considered, where the user-defined weighting $\Gb_w$ is expressed via its state-space matrices $\Ab_w,\Bb_w,\Cb_w$ and $\Db_w$. Clearly, if \eqref{eq:normalController} takes the role of the IIR filter, the error dynamics are described by
\begin{align}
\nonumber
\Ab_{e}&:=\begin{pmatrix}
\Ab_{w} & \zerob & \zerob \\
\Bb \Cb_{w} & \Ab & \zerob \\
\Bb_{f} \Cb_{w} & \zerob & \Ab_{f}
\end{pmatrix} \quad \;\;\;\;
\Bb_{e}:=\begin{pmatrix}
\Bb_{w} \\
\Bb \Db_{w} \\
\Bb_{f} \Db_{w}
\end{pmatrix}
\\
\nonumber
\Cb_{e}&\!:=\!\left(
(\Db_{f}\!-\Db) \Cb_{w} \; -\Cb \;\; \Cb_{f}
\right) \;\;
\Db_{e}\!:=\!\begin{pmatrix}\left(\Db+\Db_{f}\right) \Db\end{pmatrix}.
\end{align}
Here, the key observation is that the approximation error depends affinely on the FIR parameters $\Cb_f$ and $\Db_f$ (that have to be specified). Now, due to the well-known bounded-real lemma, the error bound $||\eb||_\infty<\gamma$ holds for some $\gamma \in \R_{>0}$ if and only if there exists a symmetric matrix $\Pb\succ 0$ of suitable dimension such that the linear matrix inequality
\begin{equation}
\label{eq:HinfCondition}
  \left(\begin{array}{ccc}
\Ab_{e}^\top \Pb \Ab_{e}-\Pb & \Ab_{e}^\top \Pb \Bb_{e} & \Cb_{e}^\top \\
\Bb_{e}^\top \Pb \Ab_{e} & -\gamma \Ib+\Bb_{e}^\top \Pb \Bb_{e} & \Db_{e}^\top \\
\Cb_{e} & \Db_{e} & -\gamma \Ib
\end{array}\right) \prec0  
\end{equation}
is feasible \cite[Thm.~1]{yamamoto}. 
We briefly note that the result of the approximation crucially depends on $\Gb_w$. A useful choice, which is recommended in \cite{yamamoto}, is a causal inverse of the IIR filter to be approximated. We further note that~\eqref{eq:HinfCondition} involves the condition  $\Ab_{e}^\top \Pb \Ab_{e}-\Pb \prec 0 $, which restricts the procedure to Schur stable $\Ab_e$ and, hence, to Schur stable $\Ab$.

\section{Benefits for encrypted control}
\label{sec:benefits}
It turns out that FIR filters have several advantages when it comes to an encrypted implementation. These will be pointed out in the following.

\subsection{Simplicity and efficiency of encryption}
\label{subsec:simplicity}
First, we focus on an integer-based reformulation of~\eqref{eq:uFIR}.
Analogously to~\eqref{eq:integerController} and~\eqref{eq:recoveries}, we easily find
\begin{equation}
\label{eq:cryptofir}
\vb_{f}(k)=\sum_{j=0}^{N} \lfloor s_6 \Fb_j \rceil  \lfloor s_7 \yb(k-j) \rceil
\end{equation}
and the recovery $\ub_f(k)\approx \vb_f(k)/(s_6 s_7)$. 
Obviously, \eqref{eq:cryptofir} does not involve any recursions. As a consequence, scaling factors are not accumulated. Moreover, $\vb_{f}(k)$ can be computed with an arithmetic circuit of multiplicative depth~$1$. These features enable simple and efficient encrypted implementations, which do not require any kind of refreshments.

Concretely, we first note that a partial encryption of~\eqref{eq:cryptofir} can be realized analogously to \cite{Farokhi2017} or \cite{Murguia2020} using, e.g., the Paillier cryptosystem. More interestingly, \eqref{eq:cryptofir} supports an efficient full encryption (i.e., of the control actions, controller parameters, and plant outputs) using leveled HE. In particular, the constant (non-accumulating) scaling factors allow choosing relatively small $q$ without suffering from overflows (of $\Z_q$). In combination with the small multiplicative depth, this allows for efficient parameter choices of the underlying LWE problem while maintaining security. We refer to \cite{albrecht2015concrete} and the corresponding online tool\footnote{\href{https://bitbucket.org/malb/lwe-estimator}{https://bitbucket.org/malb/lwe-estimator}} for details on the parameter-depending security of LWE-based leveled HE. 

\subsection{Number of operations}
\label{subsec:number}
We next compare the number of operations per time step that are required to  evaluate the dynamic controller \eqref{eq:integerController} and the FIR Filter~\eqref{eq:cryptofir}, respectively. One can easily deduce that evaluating~\eqref{eq:cryptofir} requires $lm(N+1)$ multiplications and ${m(N+l-1)}$ additions, where we note that some operations can be saved for $k<N$ if 
\begin{equation}
\label{eq:yjAssumption}
   \yb(j)=\zerob \quad \text{for every} \,\, j<0  
\end{equation}
is assumed. 
A similar analysis for \eqref{eq:integerController} is a bit more tedious, but eventually yields $(l+n)(m+n)$ multiplications and $(l+n)(m+n-1)$ additions. With these relations at hand, it can be verified that~\eqref{eq:cryptofir} requires fewer multiplications and additions than~\eqref{eq:integerController} if the inequality
\begin{equation}
\label{eq:boundForEfficientN}
\!\!N< \min \left\{ \frac{ln+mn +n^2}{lm} ,\frac{ln+n^2\!-l-n}{m}+n+1 \right\}\!
\end{equation}
holds. This inequality often leaves a lot of freedom for designing efficient FIR filters. For instance, we obtain $N<14$ for the dimensions of the example in Section~\ref{sec:benchmark}. Now, one could argue that the comparison is incomplete since only \eqref{eq:integerControllerb} has to be evaluated to obtain the current control action $\vb(k)$ while \eqref{eq:integerControllera} (and $\lfloor s_3 \Cb \rceil  \zb(k+1)$) could be computed in between the sampling instances. However, a similar observation applies to the FIR filter~\eqref{eq:cryptofir}. In fact, the right-hand side in \eqref{eq:cryptofir} can obviously be written as
$$
\lfloor s_6 \Fb_0 \rceil  \lfloor s_7 \yb(k) \rceil + \sum_{j=1}^{N} \lfloor s_6 \Fb_j \rceil  \lfloor s_7 \yb(k-j)\rceil
$$
and all but the first term can be computed before time step~$k$. Using these precomputations, both controllers require $lm$ multiplications and $lm$ additions at time step $k$.

\subsection{Used information}
\label{subsec:info}

The proposed FIR filter-based controller offers some similarities to the approach summarized in Section~\ref{subsec:reset}. In fact, among all existing encrypted dynamic controllers, only these two realizations do not require any internal or external refreshments. Furthermore, both controllers only use a finite number of past plant outputs $\yb(j)$ with $j<k$.
For the FIR filter, this number is fixed to $N$ by construction. For the reset controller, the number is  $\Delta k:=k \, \modu \, T$, and it obviously reflects the periodic resets. Using $\Delta k$, the control actions resulting from \eqref{eq:resetController} can be explicitly stated as
$$
\ub_{r}(k)= \Cb\Ab^{\Delta k} \xb_{0} + \sum_{j=0}^{\Delta k-1} \Cb \Ab^{j} \Bb \yb(k-1-j)
+\Db \yb(k)
$$
analogously to~\eqref{eq:uExact}. Now, further similarities between the reset controller \eqref{eq:resetController} and the FIR filter \eqref{eq:uFIR} are revealed for the initialization $\xb_0:=\zerob$, the  parameterization~\eqref{eq:FjWindow}, the assumption~\eqref{eq:yjAssumption}, and the special choice $N=T-1$. In fact, we then have $\ub_{r}(k)=\ub_{f}(k)$ for the first $T$ steps, i.e., for every $k \in \{0,\dots,N\}$. Moreover, the control law is identical right before every reset, i.e., for every $k$ yielding $\Delta k=N$. In all other time steps, the convolution sum associated with  $\ub_{r}(k)$ is truncated in comparison to $\ub_{f}(k)$. In other words, in these time steps, the FIR filter-based controller uses more information about  the plant than the reset controller. This should result in better control performance, as confirmed by the following numerical example.

\section{Numerical benchmark}
\label{sec:benchmark}
In order to illustrate the aforementioned benefits, we consider the same example as in \cite{Murguia2020}. 
The example encompasses the control of a chemical batch reactor. The system dynamics can be approximated based on a linear discrete-time state-space model with the matrices
\begin{align}
\nonumber
    &\Ab_{s}=\begin{pmatrix}
    \blind{+}1.18 & 0.00 & \blind{+}0.51 & -0.40 \\
    -0.05 & 0.66 & -0.01 & \blind{+}0.06\\
    \blind{+}0.08 & 0.34 & \blind{+}0.56 & \blind{+}0.38\\
    \blind{+}0.00 & 0.34 & \blind{+}0.09 & \blind{+}0.85\\
    \end{pmatrix}\;\;
    \Bb_{s}=\begin{pmatrix}
    0.00 \\
    0.47 \\
    0.21 \\
    0.21
    \end{pmatrix}
    \\
    \nonumber
    &\Cb_{s}=\begin{pmatrix}
        0 & 1 & 0 & \blind{+}0 \\ 1 & 0 & 1 & -1
    \end{pmatrix}
    \qquad \qquad \qquad \quad \;\;
    \Db_{s}=\zerob.
\end{align} 
and a sampling period of $\Delta t=0.1$ \cite[Eq.~(33)]{Murguia2020}. We briefly note that these dynamics are not trivial. In fact, the system is unstable and non-minimum phase in both outputs. Furthermore, one can show that the system cannot be stabilized based on static linear output feedback (see \cite[Sect.~IV]{Murguia2020} for details). A stabilization can, however, be achieved with a dynamic controller as in~\eqref{eq:normalController}.

In \cite{Murguia2020}, two reset controllers of the form~\eqref{eq:resetController} have been designed to enable encrypted dynamic control. The first controller, which we will denote by $\Cc_1$, is reset after $T=25$ steps, whereas the second controller $\Cc_2$ considers $T=8$. The two controllers (that are specified in \cite[{Eqs.~(34)--(35)}]{Murguia2020}) not only differ significantly in terms of $T$. In fact, the controller state matrix $\Ab$ is Schur stable for $\Cc_2$, while it is unstable for $\Cc_1$. As a consequence, the two methods for the FIR filter-based controller approximation in Section~\ref{sec:FIR} can only be applied to $\Cc_2$. However, we show that FIR filters exist that are able to compete with the (better) performance of $\Cc_1$.

\begin{figure}[tp]
	\centering
	\includegraphics[trim=3cm 11cm 3.5cm 11.2cm, clip=true,width=\linewidth]{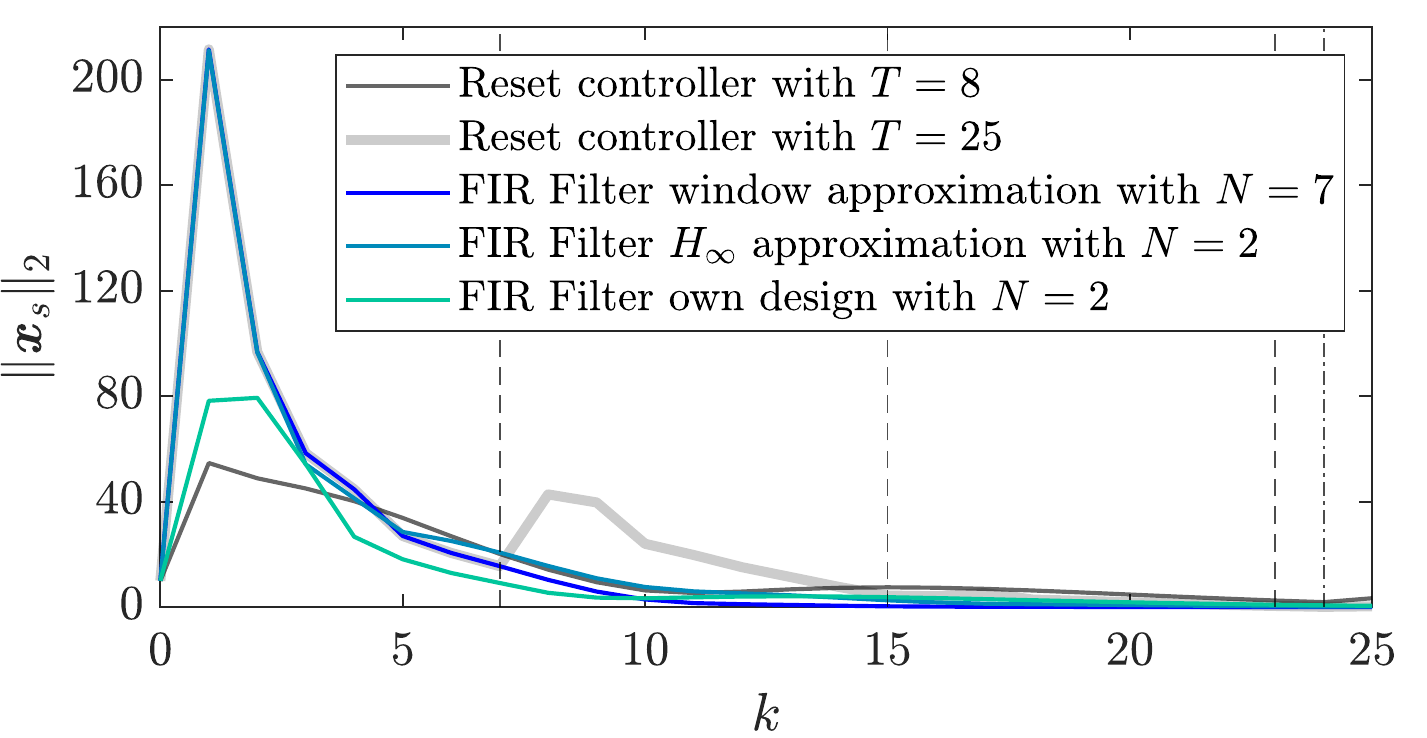}
	\caption{$2$-Norm of the system's state controlled by reset controllers and FIR filters for the initial state  $\xb_s(0)=-\left(6.83,5.18,4.05,3.12\right)^\top$. Dashed and dash-dotted lines mark the resets for $T=8$ and $T=25$, respectively.}
	\label{fig:normsys}
	\vspace{-4mm}
\end{figure}

Before we design our FIR filter-based encrypted controllers, we note that the dimensions of both $\Cc_1$ and $\Cc_2$ are $l=2$, $m=1$, and $n=4$. Hence, according to~\eqref{eq:boundForEfficientN}, a FIR filter requires fewer operations than the original IIR controllers if the filter order $N$ is chosen smaller than $\min\{14,23\}=14$. Now, inspired by the observations in Section~\ref{subsec:info}, we first design a window-based FIR filter that approximates $\Cc_2$ for the choice $N=T-1=7$ and obtain the filter matrices
\begin{align*}
\begin{pmatrix}\Fb_0 \\ \Fb_1 \\ \Fb_2 \\ \Fb_3 \end{pmatrix}\!=\!\begin{pmatrix}-49.00 & -2.33\\ \blind{+}50.99 & \blind{+}0.17\\ \,\,\,-7.31 & \blind{+}0.04\\ \,\,\,-2.42 & -0.02
\end{pmatrix}\!,
&
\,\,\begin{pmatrix} \Fb_4 \\ \Fb_5 \\ \Fb_6 \\ \Fb_7 \end{pmatrix}\!=\!\begin{pmatrix} \blind{+}0.88 & 0.00\\ \blind{+}0.03 & 0.00\\ -0.07 & 0.00\\ \blind{+}0.01 & 0.00 
\end{pmatrix}\!.
\end{align*}
Interestingly, (almost) the same matrices result for the optimization-based approach using $\gamma=0.1$ and a causal inverse of $\mathcal{C}_2$ for $\Gb_w$. This observation can be easily explained. In fact, we find here $\Cb \Ab^7 \approx \zerob$, which implies that the window-based method is nearly optimal for ${N=7}$ (cf.~Sect.~\ref{subsec:window}). However, using the optimization-based approach, we can construct a satisfactory approximation of $\mathcal{C}_2$ with a significantly smaller order. Indeed, by using $\gamma=0.06$ and the same $\Gb_w$ as above but $N=2$, we find
\begin{equation*}
    \begin{pmatrix}\Fb_0 \\ \Fb_1 \\ \Fb_2 \end{pmatrix}=
\begin{pmatrix}
-48.93 & -2.33\\ \blind{+}50.93 & \blind{+}0.17\\ \,\,\, -8.81 & \blind{+}0.04
\end{pmatrix}.
\end{equation*}
Finally, we note that the unstable controller $\Cc_1$ can likewise be replaced with a low-order FIR filter of similar performance. A suitable choice is, e.g.,
\begin{equation*}
    \begin{pmatrix}\Fb_0 \\ \Fb_1 \\ \Fb_2 \end{pmatrix}=
\begin{pmatrix}
-17.54 & -3.04\\ \,\,\, -4.44 & -0.96\\ \blind{+}17.60 & -0.23
\end{pmatrix}.
\end{equation*}
Before addressing the encryption of the designed FIR filters, we briefly investigate their control performance. As apparent from Figure~\ref{fig:normsys}, all controllers stabilize the closed-loop system. 
Moreover, $\Cc_2$ and its window-based FIR approximation are indeed identical for the first $T=8$ steps. Finally, we observe the unfavorable effects of the resets, which are not required for the FIR filters.

Now, as pointed out in Section~\ref{subsec:simplicity}, one advantage of FIR filters is that they can be fully encrypted in an efficient fashion. To confirm this feature, we implemented the window-based FIR filter with $N=7$ (i.e., the most complex one of our benchmark) using the leveled homomorphic BFV scheme introduced in~\cite{fan2012somewhat} from the PALISADE library~\cite{palisade}. We considered a ring dimension of $256$, one level, and $q=2^{20}$. With these parameters, we achieved real-time capability, i.e., computation times per time step below the sampling period of $\unit[100]{ms}$ (on an Intel Xeon E5-2620). Furthermore, we obtain a sufficient security level (of approximately $80$ bits) according to the estimator from~\cite{albrecht2015concrete}.

\section{Conclusion and Outlook}
\label{sec:outlook}
In this paper, we showed that FIR filter-based approximations of dynamic controllers are well-suited for an encrypted implementation. First, they do not suffer from limited operation times as direct realizations of dynamic controllers do. Second, they come with several additional benefits, such as a small multiplicative depth. We illustrated their advantages with a numerical example. Here, the simplicity of the FIR filter even allowed for a real-time capable fully homomorphic implementation. Taking into account that we only considered two simple FIR approximations in this paper, future research has to address more complex FIR filter-based controller designs and approximations. 

\section*{Acknowledgment}
Support by the German Research Foundation (DFG) under the grant SCHU 2940/4-1 is gratefully acknowledged.


\end{document}